\documentclass[aps,prl,reprint,showpacs,byrevtex]{revtex4-2}
\let\oldhat\hat
\renewcommand{\hat}[1]{\oldhat{\mathbf{#1}}}
\usepackage{titlesec}
\usepackage{array}

\usepackage{times,mathptm}
\usepackage[dvips]{graphicx,color}

\begin{document}

\title{Phonon-Mediated ${\bf S}$-Wave Superconductivity in the Kagome Metal CsV$_3$Sb$_5$ under Pressure}
\author{Chongze Wang$^{1,2}$, Jia Yu$^{1,3}$, Zhenyu Zhang$^4$, and Jun-Hyung Cho$^{2*}$}
\affiliation{$^1$Joint Center for Theoretical Physics, School of Physics and Electronics, Henan University, Kaifeng 475004, People's Republic of China \\
$^2$Department of Physics and Research Institute for Natural Science, Hanyang University, 222 Wangsimni-ro, Seongdong-Ku, Seoul 04763, Republic of Korea \\
$^3$Key Laboratory for Special Functional Materials of the Ministry of Education, Henan University, Kaifeng 475004, People's Republic of China \\
$^4$International Center for Quantum Design of Functional Materials (ICQD), Hefei National Laboratory for Physical Sciences at Microscale, and Synergetic Innovation Center of Quantum Information and Quantum Physics, University of Science and Technology of China, Hefei 230026, China }
\date{\today}

\begin{abstract}
The nature of the superconducting pairing state in the pristine phase of a compressed kagome metal CsV$_3$Sb$_5$ under pressure is studied by the Migdal-Eliashberg formalism and density-functional theory calculations. We find that the superconducting gap distribution driven by electron-phonon coupling is nodeless and anisotropic. It is revealed that the hybridized V 3$d$ and Sb 5$p$ orbitals are strongly coupled to the V-V bond-stretching and V-Sb bond-bending phonon modes, giving rise to a wide spread of superconducting gap depending on its associated Fermi-surface sheets and momentum. Specifically, the superconducting gaps associated with V 3$d_{xy,x^2-y^2,z^2}$ and 3$d_{xz,yz}$ orbitals are larger in their average magnitude and more widely spread compared to that associated with the Sb 5$p_z$ orbital. Our findings demonstrate that the superconductivity of compressed CsV$_3$Sb$_5$ can be explained by the anisotropic multiband pairing mechanism with conventional phonon-mediated $s$-wave symmetry, evidenced by recent experimental observations under pressure as well as at ambient pressure.
\end{abstract}
\maketitle

The recently discovered kagome metal series AV$_3$Sb$_5$ (A = K, Rb, Cs) has attracted tremendous attention due to its exotic electronic properties such as topologically nontrivial band structures, chiral charge density wave (CDW), and superconductivity (SC)~\cite{AV3Sb5-PRM2019,CsV3Sb5-Z2-PRL2020, KV3Sb5-chrialCDW-Nat.Mat2021,CsV3Sb5-Z2-PRL2020, KV3Sb5-chrialCDW-Nat.Mat2021, KV3Sb5_SC_Z2_PRM2021, RbV3Sb5-SC-CPL2021}. For CsV$_3$Sb$_5$, a charge density wave (CDW) transition occurs around 94 K at ambient pressure, followed by an emergence of SC as the temperature decreases to ${\approx}$3 K~\cite{CsV3Sb5-Z2-PRL2020}. The competition between the CDW order and SC has been intensively studied by applying pressure~\cite{CsV3Sb5-CDW_SC-NC2021, CsV3Sb5-SC_CDW-PRL2021, CsV3Sb5_SC_PRB2021, CsV3Sb5_HighPress_CPL,CsV3Sb5_HighPress_PRL,CsV3Sb5_Nodal_SC, chongze_PRB}. The observed pressure-temperature ($P$-$T$) phase diagram shows the existence of two superconducting domes under pressure~\cite{CsV3Sb5_SC_PRB2021,CsV3Sb5_HighPress_CPL,CsV3Sb5_HighPress_PRL,CsV3Sb5_Nodal_SC}. The first superconducting dome exhibits a maximum superconducting transition temperature ($T_c$) of ${\approx}$8 K around 2 GPa~\cite{CsV3Sb5-CDW_SC-NC2021, CsV3Sb5-SC_CDW-PRL2021}, while the second one exhibits a maximum $T_c$ of ${\approx}$6 K around 45 GPa~\cite{CsV3Sb5_SC_PRB2021,CsV3Sb5_HighPress_CPL,CsV3Sb5_HighPress_PRL,CsV3Sb5_Nodal_SC}. On the other hand, the CDW order is suppressed under pressure and transforms into the pristine phase at a critical pressure of ${\approx}$2 GPa~\cite{CsV3Sb5-CDW_SC-NC2021, CsV3Sb5-SC_CDW-PRL2021}. The presence of such a quantum critical point (QCP) beneath the top of the first superconducting dome resembles the $P$-$T$ phase diagrams of many unconventional superconductors such as heavy fermions~\cite{CeCu2Si2_Science}, organics~\cite{organic_JPCM2011}, and iron pnictides~\cite{Rev-pnictide-2011,Rev-pnictide-2015}, where an antiferromagnetic QCP lies beneath the superconducting dome. Here, spin fluctuations in the vicinity of magnetically ordered phases have been considered to effectively mediate the formation of Cooper pairs~\cite{3He_RMP1997, Bag_PRB1989}. Similarly, for the kagome superconductors, electronic correlation effects at ambient pressure or CDW fluctuations around the QCP were proposed to be an essential ingredient of the superconducting pairing mechanism~\cite{AV3Sb5_rev_NatPhy,AV3Sb5_rev_JAP,Kagome_Kiesel2012,Kagome_vHs2013,Kagome_Kiesel2013, AV3SB5_Wu2021, AV3SB5_Tazai2022, CsV3Sb5_Chen2022}. Meanwhile, several first-principles calculations for CsV$_3$Sb$_5$ showed that the variation of electron-phonon coupling (EPC) around the QCP plays an important role in the formation of superconducting dome~\cite{CsV3Sb5_DFT_Si,CsV3Sb5_DFT_Zhang,chongze-PRM}, supporting a conventional phonon-mediated superconducting mechanism. Thus, whether the nature of SC in CsV$_3$Sb$_5$ is unconventional (mediated by electronic interactions) or conventional (mediated by phonons) has been controversial.

The pairing symmetry of SC in CsV$_3$Sb$_5$ has also been an issue of intense debate. The symmetry structure of Cooper pairs in a superconducting state can be manifested by the momentum dependence of superconducting gap ${\Delta}$. For example, cuprate superconductors have the nodal gap with $d$-wave pairing symmetry~\cite{Rev-cuprate-2006, Rev-cuprate-2015}, while conventional phonon-mediated superconductors have the nodeless gap with $s$-wave pairing symmetry~\cite{MgB2_Nature2002}. For CsV$_3$Sb$_5$, various experiments reached different conclusions on the superconducting pairing symmetry. Ultralow temperature thermal conductivity measurements suggested nodal SC~\cite{CsV3Sb5_Nodal_SC}, whereas magnetic penetration depth experiments using tunnel diode oscillator techniques reported nodeless SC~\cite{CsV3Sb5_nodeless_mag}. Moreover, scanning tunneling microscopy and spectroscopy (STM/STS) measurements reported different superconducting features of a nodal V-shaped superconducting gap~\cite{CsV3Sb5_Roton, CsV3Sb5_nodeless_PRX} and a nodeless s-wave superconducting gap~\cite{CsV3Sb5_Multi}.

\begin{figure}[ht]
\centering{ \includegraphics[width=8.0cm]{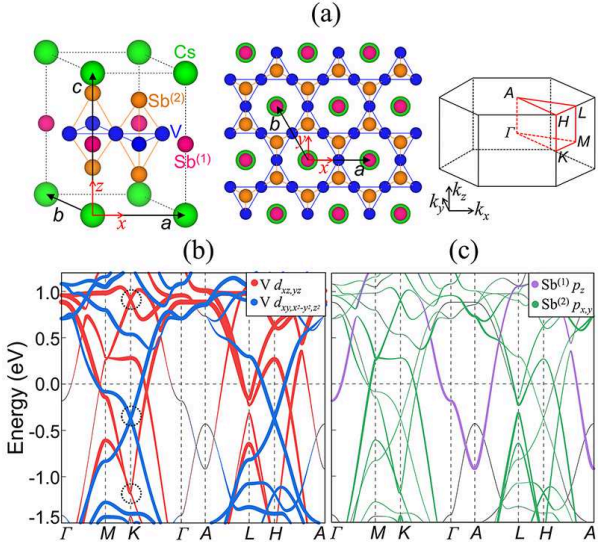} }
\caption{(a) Optimized structure of the pristine phase of CsV$_3$Sb$_5$ at 3 GPa, together with its top view (middle panel) and Brillouin zone (right panel). Here, the lattice parameters are $a$ = $b$ = 5.410 {\AA} and $c$ = 8.563 {\AA}. (b) Calculated band structure of the pristine phase at 3 GPa. The projected bands onto V 3$d$ and Sb 5$p$ orbitals are separately displayed in (b) and (c), respectively, where the radii of circles are proportional to the weights of the corresponding orbitals. For distinction, the radius scale of Sb 5$p$ orbitals is increased by two times larger compared to that of V 3$d$ orbitals.}
\end{figure}

In this Letter, using first-principles density-functional theory (DFT) calculations~\cite{DFT} together with the Migdal-Eliashberg equations~\cite{Migdal,Eliash,ME-review}, we explore the anisotropy and pairing symmetry of the superconducting gap in the pristine phase of a multiband kagome superconductor CsV$_3$Sb$_5$ under pressure. Our analysis of the band- and ${\bf k}$-resolved superconducting gap identifies the existence of nodeless, anisotropic superconducting gap distributions on the four Fermi surface (FS) sheets. Specifically, at 3 GPa and near zero temperature, the V 3$d_{xy,x^2-y^2,z^2}$, V 3$d_{xz,yz}$, and Sb 5$p_z$ orbitals give rise to a sizable gap anisotropy spreading between 1.5 and 3.5 meV, where the ${\Delta}$ associated with the former two V orbitals are larger in their average magnitude and more widely spread compared to that associated with the latter Sb one. It is revealed that the different orbitals on the multiple FS sheets are strongly coupled to the V-V bond-stretching and V-Sb bond-bending phonon modes associated with a particular V kagome net that is interwoven with Sb atoms both within and outside of the kagome plane [see Fig. 1(a)]. Interestingly, these orbital-dependent features of EPC and ${\Delta}$ in the pristine phase of compressed CsV$_3$Sb$_5$ resemble the observed~\cite{CsV3Sb5_Multi} three pairs of coherent peaks in the superocnducting gap spectra of STS at ambient pressure. The present results provide a theoretical framework for understanding the conventional phonon-mediated $s$-wave paring symmetry in the SC of the pristine phase, which was recently evidenced by experimental measurements~\cite{nanoletter-s-wave,CsV3Sb5_s-wave_NC}.

We begin by optimizing the atomic structure of CsV$_3$Sb$_5$ at a pressure of 3 GPa using the DFT scheme~\cite{methods}. Figure 1(a) shows the optimized structure corresponding to the 1${\times}$1${\times}$1 pristine phase, which crystallizes in the hexagonal space-group $P6/mmm$ (No. 191) with the stacking of the V$_3$Sb kagome layer containing a triangular Sb (termed Sb$^{(1)}$) sublattice centered on each V hexagon, the Sb (termed Sb$^{(2)}$) honeycomb layers above and below the V$_3$Sb kagome layer, and the Cs triangular layer. The electronic band structure of this pristine phase is displayed in Fig. 1(b), together with its projection onto V 3$d$ and Sb 5$p$ orbitals [see Figs. 1(b), 1(c), and S1 in the Suppelmental Material~\cite{SM}]. We find that there exist three Dirac points loacted at the $K$ point [indicated by the dashed circles in Fig. 1(b)], similar to the previous angle-resolved photoemission spectroscopy data~\cite{CsV3Sb5-ARPES-Kang, CsV3SB5-ARPES-Lou, CsV3Sb5-ARPES-Hu} measured from the high-temperature pristine phase at ambient pressure. Figures 2(a) and 2(b) show the FS composed of four sheets (designated as FS$_1$, FS$_2$, FS$_3$, and FS$_4$) at $k_z$ = 0 and ${\pi}/c$, respectively. Here, FS$_1$ forms the cylindrical-like sheet surrounding the ${\Gamma}-A$ path in the Brillouin zone [see the right panel in Fig 1(a)], while FS$_2$, FS$_3$, and FS$_4$ change their shapes due to the intertwined bands along the $k_z$ direction: i.e., FS$_2$ (FS$_3$/FS$_4$) forms the hexagonal-shaped (circular-like) sheet at $k_z$ = 0, but FS$_2$/FS$_4$ (FS$_3$) forms the circular-like (deformed hexagonal-shaped) sheet at $k_z$ = ${\pi}/c$. In Figs. 2(a) and 2(b), we display the projected FS sheets onto the V 3$d$ and Sb 5$p$ orbitals. We find that the FS sheets feature different orbital characters: i.e., FS$_1$ arises mostly from the Sb$^{(1)}$ $p_z$ orbital, FS$_2$ from the V $d_{xy,x^2-y^2,z^2}$ orbitals, and FS$_3$ and FS$_4$ from the V $d_{xz,yz}$ orbitals. It is noticeable that the V $d_{xy,x^2-y^2,z^2}$ and $d_{xz,yz}$ orbitals around $E_F$ hybridize conspicuously with the Sb$^{(2)}$ $p_{x,y}$ orbitals [see Figs. 1(b) and 1(c)]~\cite{Adv.Matearial-expt}. This hybridization between V 3$d$ and Sb 5$p$ orbitals leads to an effective electron-phonon interaction between the V$_3$Sb$^{(1)}$ kagome and Sb$^{(2)}$ honeycomb layers, as discussed below. We will also demonstrate later that the presence of such multiple FS sheets with different orbital characters provides a strong anisotropy in EPC, thereby yielding a multiband SC with highly anisotropic superconducting-gap distributions.

\begin{figure}[ht]
\centering{ \includegraphics[width=8.0cm]{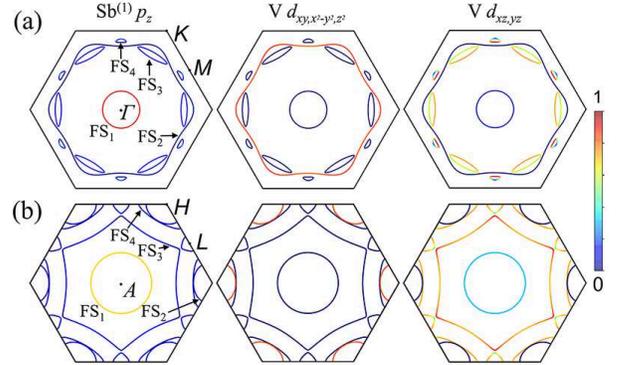} }
\caption{FS sheets of the pristine phase at 3 GPa, projected onto the Sb$^{(1)}$ 5$p_z$, V 3$d_{xy,x^2-y^2,z^2}$, and V 3$d_{xz,yz}$ orbitals at (a) $k_z$ = 0 and (b) ${\pi}/c$ using the color scale.}
\end{figure}

To explore the EPC in compressed CsV$_3$Sb$_5$, we calculate the phonon spectrum with the EPC strength of each phonon mode, Eliashberg spectral function ${\alpha}^{2}F$, and integrated EPC constant ${\lambda}({\omega})$ as a function of phonon frequency. The calculated results at 3 GPa are displayed in Figs. 3(a) and 3(b). We find that there are two frequency regimes $R_1$ and $R_2$ where ${\lambda}({\omega})$ increases as large as ${\approx}$80\% and ${\approx}$20\% of the total EPC constant ${\lambda}$ = ${\lambda}$(${\infty}$) = 1.39, respectively [see Fig. 3(b)]. It is noticeable that the phonon modes $L_1$, $L_2$, and $M_1$ in the low-frequency $R_1$ regime and $L_3$ and $M_2$ in the high-frequency $R_2$ regime exhibit large EPC strengths. As shown in the right panel of Fig. 3(a), $L_1$ and $M_1$ represent the V-V bond-stretching modes coupled with the up and down vibration of Sb$^{(2)}$ atoms; $L_2$ is similar to $L_1$ but also involves the relatively larger up and down vibration of Cs atoms approaching Sb$^{(1)}$ atoms; and $L_3$ and $M_2$ represent the V-Sb$^{(2)}$ bond-bending modes (see the animation of each mode in the Supplemental Material~\cite{SM}). As pressure increases, the $L_i$ and $M_i$ ($i$ = 1, 2, or 3) phonon modes increase their frequencies [see Fig. 3(c)], yielding a sharp decrease in ${\lambda}$ as 0.84 and 0.71 at 5 and 7 GPa, respectively [see Figs. 2(b)]. In other words, as pressure approaches a QCP of ${\approx}$2 GPa from higher pressures, the softening of the $L_i$ and $M_i$ phonon modes increases ${\lambda}$, leading to the formation of a superconducting dome around the QCP~\cite{CsV3Sb5-CDW_SC-NC2021, CsV3Sb5-SC_CDW-PRL2021, chongze-PRM}. Here, the soft $L_i$ ($M_i$) modes at the three equivalent $L$ ($M$) points induce a quantum phase transition to the 2${\times}$2${\times}$2 CDW phase with the so-called inverse-star-of-David structure~\cite{AV3Sb5-B.Yan-PRL2021}.

\begin{figure}[ht]
\centering{ \includegraphics[width=8.0cm]{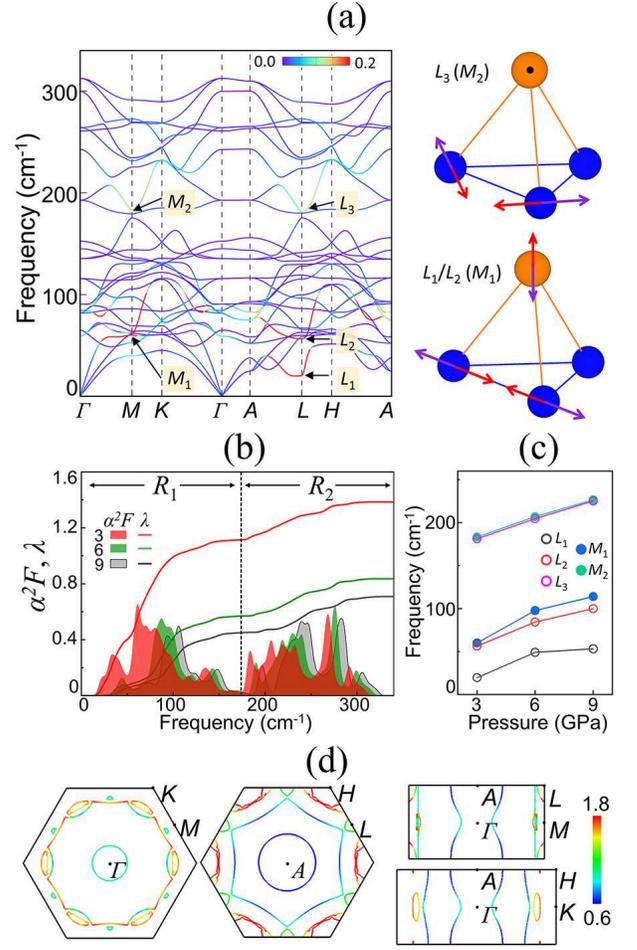} }
\caption{(a) Calculated phonon spectrum of the pristine phase at 3 GPa, together with the EPC strength (in color scale) of each phonon mode and the main atomic displacements of $L_i$ and $M_i$ ($i$ = 1, 2, or 3) modes. Here, $L_i$ and $M_i$ belong to the $B_{1u}$ and $A_g$ modes with $D_{2h}$ symmetry, respectively. The results of ${\alpha}^{2}F$ and ${\lambda}({\omega})$ obtained at 3, 5, and 7 GPa are given in (b). The frequencies of $L_i$ and $M_i$ ($i$ = 1, 2, or 3) modes as a function of pressure are displayed in (c). In (d), ${\lambda}_{n{\bf k}}$ on the FS is drawn on the horizontal $k_z$ = 0 and $k_z$ = ${\pi}/c$ planes and the vertical ${\Gamma}-M-L-A$ and ${\Gamma}-K-H-A$ planes.}
\end{figure}

Next, the anisotropy of EPC in compressed CsV$_3$Sb$_5$ is examined by using the anisotropic Migdal-Eliashberg equations~\cite{Migdal,Eliash,ME-review}. We calculate the $n$- and ${\bf k}$-resolved EPC constant ${\lambda}_{n{\bf k}}$, which includes all available electron-phonon scattering processes connecting ${\bf k}$ and other ${\bf k}$ points on the FS$_n$ ($n$ = 1, 2, 3, 4) sheets. Figure 3(d) shows ${\lambda}_{n{\bf k}}$ on the FS$_n$ sheets at 3 GPa. We find that ${\lambda}_{n{\bf k}}$ associated with the Sb$^{(1)}$ $p_z$ orbital ($n$ = 1) distributes between ${\approx}$0.6 and ${\approx}$1.0. Meanwhile, ${\lambda}_{n{\bf k}}$ associated with the V 3$d$ orbitals is quite widely spread between ${\approx}$0.7 and ${\approx}$2.2, where the V ($d_{xz,yz}$: $n$ = 3 and 4) and ($d_{xy,x^2-y^2,z^2}$: $n$ = 2) orbitals are in the ranges of 0.7$-$1.6 and 1.5$-$2.2, respectively. Therefore, the EPC strength of the electronic states on the FS sheets varies with respect to their orbital characters and ${\bf k}$ directions, indicating a strong anisotropy of EPC. It is worth noting that this orbital-dependent EPC is attributed to the specific three-dimensional bonding character of the CsV$_3$Sb$_5$ kagome crystal: i.e., (i) the V $d_{xy,x^2-y^2}$ ($d_{z^2}$) orbitals forming the V-V ${\sigma}$(${\pi}$)-bonding states are effectively coupled to the V-V bond-stretching phonon modes, (ii) the V $d_{xz,yz}$ orbitals hybridizing with the Sb$^{(2)}$ $p_{x,y}$ orbitals [see Figs. 1(b) and 1(c)] are coupled to the V-Sb$^{(2)}$ bond-bending phonon modes, and (iii) the Sb$^{(1)}$ $p_z$ orbital on the FS$_1$ sheet is coupled to the $L_2$ phonon mode that involves a decrease in the Cs$-$Sb$^{(1)}$ distance due to the up and down vibration of Cs atoms, as mentioned above.

It is natural that the wide distribution of ${\lambda}_{n{\bf k}}$ leads to an anisotropy in ${\Delta}$. By numerically solving the anisotropic Migdal-Eliashberg equations~\cite{Migdal,Eliash,ME-review} with a typical Coulomb pseudopotential parameter of ${\mu}^*$ = 0.13~\cite{Allen,MgB2_PRL2001,chongze-PRM}, we calculate the temperature dependence of ${\Delta}$ at 3 GPa. Figure 4(a) displays the energy distribution of ${\Delta}$ as a function of temperature. We find that the widely distributed ${\Delta}$ closes at a $T_{\rm c}$ of ${\approx}$15 K. To analyze the anisotropy of ${\Delta}$, we calculate the $n$- and ${\bf k}$-resolved superconducting gap ${\Delta}_{n{\bf k}}$ on the FS sheets at 2 K. As shown in Fig. 4(b), ${\Delta}_{n{\bf k}}$ associated with the Sb$^{(1)}$ $p_z$ ($n$ = 1), V $d_{xz,yz}$ ($n$ = 3 and 4), and V $d_{xy,x^2-y^2,z^2}$ ($n$ = 2) orbitals are in the ranges of 1.5$-$2.2, 1.6$-$2.5, and 2.3$-$3.5 meV, respectively. These orbital- and ${\bf k}$-dependent features of ${\Delta}_{n{\bf k}}$ are similar to the corresponding ${\lambda}_{n{\bf k}}$ ones, indicating that both ${\Delta}_{n{\bf k}}$ and ${\lambda}_{n{\bf k}}$ are correlated with each other. It is noticeable that ${\Delta}_{n{\bf k}}$ on each FS sheet is spread without any node, which represents an anisotropic superconducting gap with s-wave pairing symmetry. In Fig. 4(a), the dashed line represents the ${\Delta}$ vs $T$ curve obtained using the isotropic Migdal-Eliashberg formalism~\cite{epw-iso}. Here, we obtain $T_{\rm c}$ ${\approx}$13 K, slightly lower than that (${\approx}$15 K) estimated using the anisotropic Migdal-Eliashberg formalism. Note that these theoretical $T_c$ values based on the harmonic approximation are overestimated compared to the experimental data of ${\approx}$8 K at the QCP~\cite{CsV3Sb5-CDW_SC-NC2021, CsV3Sb5-SC_CDW-PRL2021}. However, the dimensionless ratio 2${\Delta}_{T=0}$/$k_B$$T_c$ with the isotropic gap and $T_c$ is 4.28 at 3 GPa, well comparable with the experimental~\cite{nanoletter-s-wave} values of 5.20 and 4.66 at 2.87 and 3.99 GPa, respectively. These larger theoretical and experimental ratios than the weak-coupling BCS value of 3.52 indicate a strong-coupling SC in CsV$_3$Sb$_5$.

\begin{figure}[htb]
\centering{ \includegraphics[width=8.0cm]{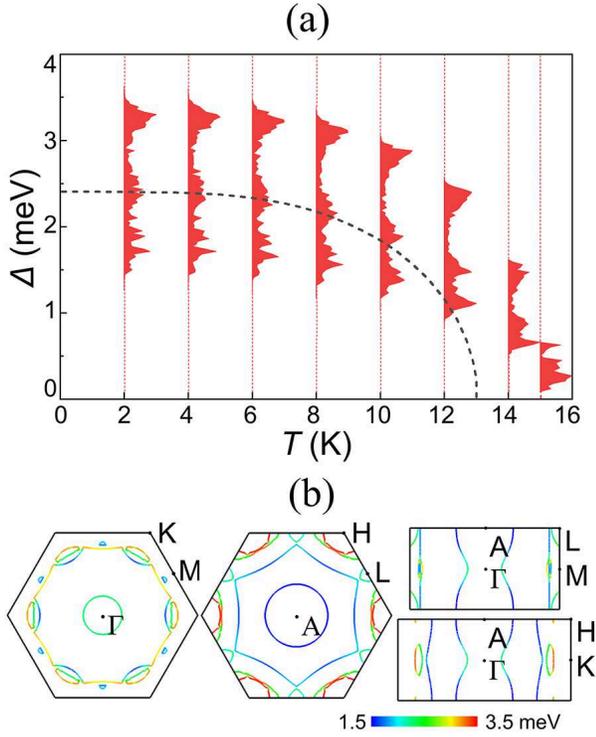} }
\caption{(a) Calculated energy distribution of the anisotropic superconducting gap as a function of temperature at 3 GPa and (b) ${\Delta}_{n{\bf k}}$ on the FS at 2 K. The dashed line in (a) represents ${\Delta}$ values, estimated using the isotropic Migdal-Eliashberg formalism. }
\end{figure}

To examine how the characteristics of anisotropic SC vary with increasing pressure, we calculate ${\lambda}_{n{\bf k}}$ and ${\Delta}_{n{\bf k}}$ at 5 GPa. The calculated distribution of ${\lambda}_{n{\bf k}}$ on the FS is displayed in Fig. S2(a)~\cite{SM}. We find that ${\lambda}_{n{\bf k}}$ associated with the Sb$^{(1)}$ $p_z$ ($n$ = 1), V $d_{xz,yz}$ ($n$ = 3 and 4), and V $d_{xy,x^2-y^2,z^2}$ ($n$ = 2) orbitals distribute in the ranges of 0.5$-$0.9, 0.6$-$1.0, and 0.8$-$1.2, respectively. Here, the magnitude and distribution of ${\lambda}_{n{\bf k}}$ arising from V $d$ orbitals are much reduced compared to the corresponding ones at 3 GPa, but the orbital-dependent features of ${\lambda}_{n{\bf k}}$ are similar between 3 and 5 GPa. Due to the reduced ${\lambda}_{n{\bf k}}$ values at 5 GPa, the temperature dependence of ${\Delta}$ closes at $T_c$ = 9 K (see Fig. S3) with 2${\Delta}_{T=0}$/$k_B$$T_c$ ${\approx}$ 3.83. Therefore, as pressure increases, 2${\Delta}_{T=0}$/$k_B$$T_c$ is lowered towards the BCS weak-coupling limit. As shown in Fig. S2(b), the ${\Delta}_{n{\bf k}}$ distributions on the FS at 2 K are in the ranges of 1.0$-$1.7, 1.1$-$1.6, and 1.4$-$2.2 meV for the Sb$^{(1)}$ $p_z$ ($n$ = 1), V $d_{xz,yz}$ ($n$ = 3 and 4), and V $d_{xy,x^2-y^2,z^2}$ ($n$ = 2) orbitals, respectively. It is thus likely that the pristine phase at higher pressures preserves the anisotropic superconducting characteristics with $s$-wave pairing symmetry.

Recently, a high-resolution STM/STS experiment~\cite{CsV3Sb5_Multi} measured the superconducing-gap spectra from a honeycomb Sb surface at ambient pressure, where three pairs of coherent peaks are located at ${\Delta}_1$ = 0.48 meV, ${\Delta}_2$ = 0.36 meV, and ${\Delta}_3$ = 0.38 meV. Although these observed gap sizes in the CDW phase are much smaller than the orbital-dependent gap sizes presently predicted from the pristine phase at 3 or 5 GPa, their nature of ${\Delta}$ seems to be similar with each other. According to STM/STS measurements~\cite{CsV3Sb5_Multi}, ${\Delta}_1$ and ${\Delta}_2$ are anisotropic on the FS sheets that are related to the CDW formation in V kagome lattice, while ${\Delta}_3$ is isotropic on the FS sheet that plays a minor role in the CDW. Indeed, the present results for the orbital dependence of ${\Delta}_{n{\bf k}}$ show that the V 3$d_{xz,yz}$ and 3$d_{xy,x^2-y^2,z^2}$ orbitals consisting of the FS$_2$, FS$_3$, and FS$_4$ sheets contribute to a wider spread of ${\Delta}$ compared to the Sb$^{(1)}$ 5$p_z$ orbital consisting of the FS$_1$ sheet. It is remarkable that STM/STS measurements~\cite{CsV3Sb5_Multi} claimed the anisotropic multiband SC with s-wave pairing symmetry in the CDW phase at ambient pressure, analogous to that presently predicted from the compressed pristine phase. Furthermore, both transport/self-field critical current experiments~\cite{nanoletter-s-wave} and electron irradiation/magnetic penetration depth measurements~\cite{CsV3Sb5_s-wave_NC} provide strong evidence for the presence of a conventional $s$-wave SC in the CDW phase at ambient pressure as well as the pristine phase under pressure.

In summary, our first-principles calculations for the pristine phase of compressed CsV$_3$Sb$_5$ have shown that the V 3$d_{xy,x^2-y^2,z^2}$, V 3$d_{xz,yz}$, and Sb$^{(1)}$ 5$p_z$ orbitals on the multiple FS sheets are strongly coupleed to the V-V bond-stretching and V-Sb bond-bending phonon modes, giving rise to the orbital- and momentum-dependent distributions of ${\lambda}_{n{\bf k}}$ and ${\Delta}_{n{\bf k}}$. Therefore, we explain the observed SC of the pristine phase in terms of a highly anisotropic multiband SC with the phonon-mediated $s$-wave pairing mechanism. Our findings have important implications for understanding the nature of the superconducting pairing state in a new family of kagome superconductors AV$_3$Sb$_5$ (A = K, Rb, Cs).

\noindent {\bf Acknowledgements.}
This work was supported by the National Research Foundation of Korea (NRF) grant funded by the Korean Government (Grant No. 2022R1A2C1005456) and by BrainLink program funded by the Ministry of Science and ICT through the National Research Foundation of Korea (Grant No. 2022H1D3A3A01077468). The calculations were performed by the KISTI Supercomputing Center through the Strategic Support Program (Program No. KSC-2022-CRE-0073) for the supercomputing application research.  \\

\noindent $^{*}$ Corresponding author: chojh@hanyang.ac.kr

\end{document}